\documentclass[letterpaper, 10 pt, conference]{ieeeconf}  

\IEEEoverridecommandlockouts

\let\proof\relax 
\let\endproof\relax 

\usepackage{bbm}
\usepackage{mathptmx}
\usepackage{graphicx} 
\usepackage{amssymb}
\usepackage{url}
\usepackage{subfigure}
\usepackage{amsmath}
\usepackage{amsthm,amsfonts,bm}
\usepackage{soul}
\usepackage{cite}
\usepackage{hyperref}
\usepackage{algorithm,algpseudocode}

\usepackage{tikz}
\usetikzlibrary{arrows}

\newtheorem{assumption}{Assumption}

\newtheorem{proposition}{Proposition}

\begin{document}

\title{\huge{Dynamic and Distributed Optimization for the Allocation of Aerial Swarm Vehicles}
}

\author{Jason Hughes, Dominic Larkin, Charles O'Donnell, Christopher Korpela
\thanks{The authors are with the Robotics Research Center, Department of Electrical Engineering and Computer Science, United States Military Academy, West Point, NY, 10996 USA. E-mail: \{jason.hughes, dominic.larkin, charles.o'donnell, christopher.korpela\}@westpoint.edu }}

\maketitle

\begin{abstract}
Optimal transport (OT) is a framework that can guide the design of efficient resource allocation strategies in a network of multiple sources and targets. This paper applies discrete OT to a swarm of UAVs in a novel way to achieve appropriate task allocation and execution. Drone swarm deployments already operate in multiple domains where sensors are used to gain knowledge of an environment \cite{sanders2017}. Use cases such as, chemical and radiation detection, and thermal and RGB imaging create a specific need for an algorithm that considers parameters on both the UAV and waypoint side and allows for updating the matching scheme as the swarm gains information from the environment. Additionally, the need for a centralized planner can be removed by using a distributed algorithm that can dynamically update based on changes in the swarm network or parameters. To this end, we develop a dynamic and distributed OT algorithm that matches a UAV to the optimal waypoint based on one parameter at the UAV and another parameter at the waypoint. We show the convergence and allocation of the algorithm through a case study and test the algorithm's effectiveness against a greedy assignment algorithm in simulation. 
\end{abstract}


\section{Introduction}
Optimal transport (OT) is a centralized framework that is often leveraged for resource allocation from a set of source nodes to a set of target nodes \cite{galichon2018optimal}. OT has been used for many efficient resource allocation and matching problems such as allocating raw materials for consumption, matching employees to tasks in a corporate environment, and allocating limited power units in areas struck by natural disasters.

While optimal transport is used for many problems, it has yet to be adapted for mapping members of a UAV swarm to a series of waypoints in a dynamic and distributed manner. To this end, the main contribution of this paper is a dynamic and distributed optimal transport algorithm for the efficient mapping of UAVs to waypoints. We consider this research as an initial development that will lead to more complex matching problems with a heterogeneous swarm. 

\begin{figure}[t]
	\centering
	\includegraphics[width=1\linewidth]{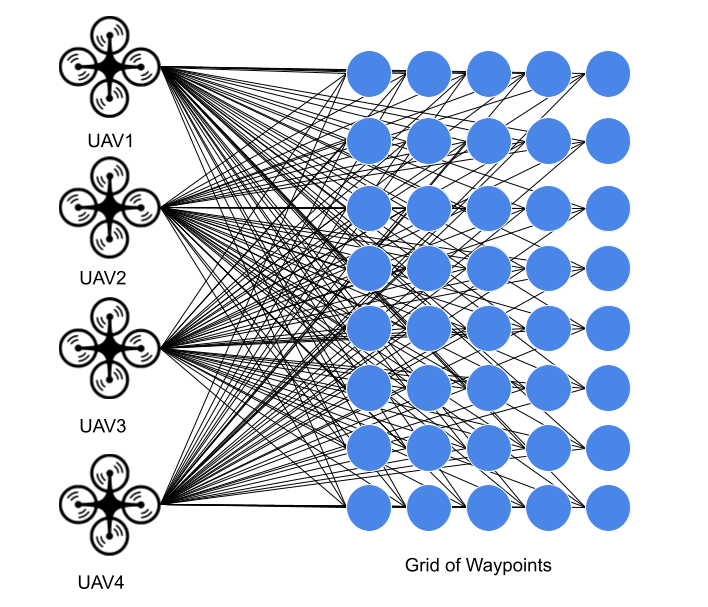}
	\caption{A Network $\mathcal{G}$ with four UAV nodes and forty waypoint nodes, and a full connection scheme where every UAV is connected to every waypoint node.}
	\label{fig:intro_image}
\end{figure}

Leveraging the discrete optimal transport formulation has many advantages. First, consider a network where both the UAV and the waypoint each have a parameter to be accounted for in the optimization algorithm, as shown in Fig. \ref{fig:intro_image}. In this network, the distance between the UAV and the waypoint is the agent side parameter, and the importance of visiting a waypoint is the waypoint side parameter. With the OT algorithm, these parameters exist at the connection between the nodes rather than at the nodes. Secondly, a dynamic algorithm is formed to update the matching scheme when parameters in the network change. The network parameters only change when a UAV visits a waypoint. This constraint avoids the possibility of visiting the same waypoint twice and prevents the algorithm from becoming stuck in a local minimum. The algorithm also must account for when a UAV has to land for a  battery swap. Finally, consider an algorithm for application to a task where information is gained only at a waypoint. Examples of these tasks include chemical sensing, radiation sensing, or surveying an area.  

A problem to be considered is as the network becomes larger and larger with more nodes, the computational complexity grows exponentially. As illustrated in Fig \ref{fig:intro_image}, there is a relatively small number of waypoints (40) and four UAVs, but there are 120 connections resulting in 240 parameters. This problem becomes more profound when the allocation occurs sequentially over time, as it does for our matching problem. In this case, parameters are added at each iteration, adding complexity. Since centralized control of the swarm is undesirable, we developed a distributed algorithm using the Alternating Direction Method of Multipliers (ADMM). This algorithm allows each UAV to solve its own optimization problem and then communicate its results with the other swarm members. 

Another consideration is that the network changes as time progresses. This change could lead to the undesirable behavior of swarm members visiting the same waypoint more than once. There is also the problem of battery life. At some point, UAVs may need to drop out of the swarm for a battery swap or a malfunction. Lastly, as a UAV visits a waypoint and takes a sensor reading, information is gained about the waypoint side of the network. The network parameters change with this additional information, and a new optimal matching now exists. These requirements lead to developing a dynamic distributed algorithm that calculates an optimal solution any time the parameters and network structure update.

Furthermore, consider the information that needs exchanging at each step. The UAVs communicate by sending Wi-Fi packets when the algorithm runs on hardware. This creates a need to keep the amount of data being transferred as small as possible to support the scalability of the swarm. Once a waypoint is reached, the proposed algorithm only needs to share three integers with the other agents in the swarm, thus limiting the amount of data transferred. 

The contributions of this paper are summarized as follows. First, a centralized optimal transport algorithm is leveraged for the UAV to waypoints matching problem. Second, a distributed algorithm is formed using ADMM, so there is no need for a centralized control station. Third, a dynamic distributed algorithm is created that can automatically calculate the optimal solution to the matching problem when parameters are updated. Lastly, we corroborate our results with a preliminary study in MATLAB and show the algorithm's effectiveness in simulation. 

\textit{Related Work.} Swarming UAVs have been widely studied \cite{arnold2019robot}, \cite{arnold2021performance}, \cite{brick2018}, \cite{sahin2005swarms}. The centralized and distributed optimal transport algorithm that we leverage is developed in \cite{zhang2019consensus}. A dynamic and distributed optimal transport algorithm is shown in \cite{jhughes2021fair}, and a secure distributed OT algorithm is developed in \cite{jhughes2021fair}. Optimal transport has been used for swarm guidance in \cite{behcet2014probabilisticOT}, \cite{vishaal2018dot} but has yet to be used dynamically. A dynamic OT algorithm is developed in \cite{wang2020dynamic} but it is not distributed. Radioactivity sensing swarms refer to waypoint allocation but do not use an OT based algorithm \cite{savidge2019radiationswarm},\cite{kopeiken2019swarmdata}. Distributed resource matching and allocation algorithms are studied extensively in \cite{Ghorbanzadeh2017}, \cite{Niu2013distributedRA}, \cite{Schmidt2009distrivutedSchemes} and specifically optimal transport in \cite{galichon2018optimal}. 



\section{Algorithm Formulation}
In this section we set up the discrete optimal transport problem with the UAV to waypoint matching problem. 

\subsection{Discrete Optimal Transport}
Start by considering each UAV as a "target" node represented by $x\in\mathcal{X}$ and each waypoint as a "source" represented by $y \in \mathcal{Y}$ and define $\mathcal{N}=\{\mathcal{X}+\mathcal{Y}\}$ as the set of all nodes. Each source node is connected to some or all target nodes, the set of targets that are connected to source $y$ is denoted by $\mathcal{X}_y$, alternatively the set of source nodes connected to target $x$ is denoted by $\mathcal{Y}_x$. For convenience, we denote the set of edges connecting target and source nodes as $\mathcal{E}:=\{\{x,y\}|x\in\mathcal{X}_y,y\in\mathcal{Y}_x\}$. We denote the network made up of the nodes and edges as $\mathcal{G} = \{\mathcal{N},\mathcal{E}\}$. With this notation, we use the optimal transport formulation from \cite{zhang2019consensus} that considers utility from both the target and source sides. We have the following discrete optimal transport formulation:
\begin{equation}\label{eqn:centralOT1}
\begin{aligned}
    \max_{\Pi}\ \sum_{x\in\mathcal{X}} \sum_{y\in\mathcal{Y}_x}& d_{xy}(\pi_{xy}) + \sum_{y\in\mathcal{Y}} \sum_{x\in\mathcal{X}_y} s_{xy}(\pi_{xy})\\
    \mathrm{s.t.}\quad &\underline{p}_{x}\leq \sum_{y\in\mathcal{Y}_x} \pi_{xy}\leq \bar{p}_{x},\ \forall x\in\mathcal{X},\\
    &\underline{q}_{y}\leq \sum_{x\in\mathcal{X}_y} \pi_{xy}\leq \bar{q}_{y},\ \forall y\in\mathcal{Y},\\
    &\pi_{xy}\geq 0,\ \forall \{x,y\} \in\mathcal{E},
\end{aligned}
\end{equation}
where $d_{xy}:\mathbb{R}_+\rightarrow\mathbb{R}$ and $s_{xy}:\mathbb{R}_+\rightarrow\mathbb{R}$ are utility functions for target node $x$ and source node $y$, respectively. They capture chosen utilities at the UAV and target side. The upper and lower bounds are shown by $\bar{p}_x$, $\bar{q}_y$ and $\underline{p}_x$, $\underline{q}_y$, respectively. These bounds mean that target $x$ will not take in more than $\bar{p}_x$ and take in less than $\underline{p}_x$. It follows similarly for the source nodes.

\subsection{Distributed Discrete Optimal Transport }
To form a distributed algorithm such that each UAV can solve its own problem, we first make adjustments to the variables. First, the ancillary variables $\pi_{xy,d}$ and $\pi_{xy,s}$ are introduced, where subscripts $d$ and $s$ indicate the relation to destination or source nodes, respectively. We then set $\pi_{xy} = \pi_{xy,d}$ and $\pi_{xy,s} = \pi_{xy}$, doing so indicates that the proposed solutions by the target and source nodes are consistent.Then, the opposite of the original functions is taken to make a convex optimization problem under the following assumption:
\begin{assumption}\label{assump1}
The utility functions $d_{xy}$ and $s_{xy}$ must be concave and monotonically increasing. 
\end{assumption}
With this assumption and the introduction of ancillary variables, we can reformulate \eqref{eqn:centralOT1} to the following:
\begin{equation}\label{eqn:centralOT2}
\begin{aligned}
\min_{\Pi_t \in \mathcal{F}_t, \Pi_s \in \mathcal{F}_s,\Pi} & -\sum_{x\in\mathcal{X}} \sum_{y\in\mathcal{Y}_x} d_{xy}(\pi_{xy,d}) - \sum_{y\in\mathcal{Y}} \sum_{x\in\mathcal{X}_y} s_{xy}(\pi_{xy,s}) \\
\mathrm{s.t.}\quad & \pi_{xy,d} = \pi_{xy},\ \forall \{x,y\}\in\mathcal{E},\\
& \pi_{xy,s} = \pi_{xy},\ \forall \{x,y\}\in\mathcal{E},
\end{aligned}
\end{equation}
where $\Pi_t:=\{\pi_{xy,d}\}_{x\in\mathcal{X}_y,y\in\mathcal{Y}}$, $\Pi_s:=\{\pi_{xy,s}\}_{x\in\mathcal{X},y\in\mathcal{Y}_x}$, $\mathcal{F}_t := \{ \Pi_t | \pi_{xy,d} \geq 0, \underline{p}_x \leq \sum_{y \in \mathcal{Y}_x} \pi_{xy,d} \leq \bar{p}_x,\ \{x,y\} \in \mathcal{E}\}$, and $\mathcal{F}_s := \{ \Pi_s | \pi_{xy,s} \geq 0, \underline{q}_y \leq \sum_{x \in \mathcal{X}_y} \pi_{xy,s} \leq \bar{q}_y,\ \{x,y\} \in \mathcal{E} \}$.

With the convex optimization problem in \eqref{eqn:centralOT2} we can apply the Alternating Direction Method of Multipliers (ADMM) to form a distributed algorithm \cite{boyd2011distributed}. 

We first form a Lagrangian with $\alpha_{xy,d}$ and $\alpha_{xy,s}$ as Lagrangian multipliers for the constraints  $\pi_{xy,d} = \pi_{xy}$ and $\pi_{xy} = \pi_{xy,s}$.
\begin{equation}\label{eqn:lagr}
\begin{split}
    &L \left(\Pi_{t}, \Pi_{s}, \Pi, \alpha_{xy,d}, \alpha_{xy,s} \right) =- \sum_{x\in\mathcal{X}} \sum_{y\in\mathcal{Y}_x} d_{xy}(\pi_{xy,d})\\
    & - \sum_{y\in\mathcal{Y}} \sum_{x\in\mathcal{X}_y} s_{xy}(\pi_{xy,s})  +\sum_{x\in\mathcal{X}} \sum_{y\in\mathcal{Y}_x} \alpha_{xy,d} (\pi_{xy,d} - \pi_{xy}) \\ &+\sum_{y\in\mathcal{Y}} \sum_{x\in\mathcal{X}_y} \alpha_{xy,s} (\pi_{xy} - \pi_{xy,s}) + \frac{\eta}{2} \sum_{x\in\mathcal{X}} \sum_{y\in\mathcal{Y}_x} (\pi_{xy,d} - \pi_{xy})^2 \\
    &+ \frac{\eta}{2} \sum_{y\in\mathcal{Y}} \sum_{x\in\mathcal{X}_y} (\pi_{xy} - \pi_{xy,s})^2,
\end{split}
\end{equation}
where $\eta > 0$ is a positive scalar controlling the convergence rate of the following algorithm. Note that in \eqref{eqn:lagr}, the last two terms $\frac{\eta}{2} \sum_{x\in\mathcal{X}} \sum_{y\in\mathcal{Y}_x} (\pi_{xy,d} - \pi_{xy})^2$ and $\frac{\eta}{2} \sum_{y\in\mathcal{Y}} \sum_{x\in\mathcal{X}_y} (\pi_{xy} - \pi_{xy,s})^2$, acting as penalization, are quadratic. Hence, the Lagrangian function $L$ is strictly convex, ensuring the existence of a unique optimal solution.

Next, we apply ADMM to the minimization problem in \eqref{eqn:centralOT2} with the Lagrangian to form the distributed algorithm:
\begin{proposition}
 The simplified iterative steps of applying ADMM to problem \eqref{eqn:centralOT2} are summarized as follows, the simplification comes from \cite{zhang2019consensus}:
\begin{equation}\label{ADMM2_eqn1}
\begin{split}
    \Pi_{x,d}(k+1) &\in \arg \min_{\Pi_{x,d}\in\mathcal{F}_{x,d}} - \sum_{y\in\mathcal{Y}_x} d_{xy}(\pi_{xy,d}) \\
   & + \sum_{y\in\mathcal{Y}_x} \alpha_{xy}(k) \pi_{xy,d} + \frac{\eta}{2} \sum_{y\in\mathcal{Y}_x} \left(\pi_{xy,d} - \pi_{xy}(k)\right)^2,
\end{split}
\end{equation}
\begin{equation}\label{ADMM2_eqn2}
\begin{split}
    \Pi_{y,s}(k+&1) \in \arg \min_{\Pi_{y,s}\in\mathcal{F}_{y,s}} - \sum_{x\in\mathcal{X}_y} s_{xy}(\pi_{xy,s}) \\ &-\sum_{x\in\mathcal{X}_y} \alpha_{xy}(k)\pi_{xy,s} + \frac{\eta}{2} \sum_{x\in\mathcal{X}_y} \left(\pi_{xy}(k) - \pi_{xy,s}\right)^2,
\end{split}
\end{equation}
\begin{equation}\label{ADMM2_eqn3}
\begin{split}
    \pi_{xy}(k+1) = \frac{1}{2} \left(\pi_{xy,d}(k+1) + \pi_{xy,s}(k+1)\right),
\end{split}
\end{equation}
\begin{equation}\label{ADMM2_eqn4}
\begin{split}
    \alpha_{xy}(k+1) = \alpha_{xy}(k) + \frac{\eta}{2}\left(\pi_{xy,d}(k+1) - \pi_{xy,s}(k+1)\right),
\end{split}
\end{equation}
where $\Pi_{\tilde{x},t}:=\{\pi_{xy,d}\}_{y\in\mathcal{Y}_x,x=\tilde{x}}$ represents the solution at target node $\tilde{x}\in\mathcal{X}$, and $\Pi_{\tilde{y},s}:=\{\pi_{xy,s}\}_{x\in\mathcal{X}_y,y=\tilde{y}}$ represents the proposed solution at source node $\tilde{y}\in\mathcal{Y}$. In addition, $\mathcal{F}_{x,d} := \{ \Pi_{x,d} | \pi_{xy,d} \geq 0, y\in\mathcal{Y}_x, \underline{p}_x \leq \sum_{y \in \mathcal{Y}_x} \pi_{xy,d} \leq \bar{p}_x\}$, and $\mathcal{F}_{y,s} := \{ \Pi_{y,s} | \pi_{xy,s} \geq 0, x\in\mathcal{X}_y, \underline{q}_y \leq \sum_{x \in \mathcal{X}_y} \pi_{xy,s} \leq \bar{q}_y\}$.
\end{proposition}
For convenience, we provide these steps in the form of an algorithm:
\begin{algorithm}[!h]
\caption{Distributed OT Algorithm}\label{Alg:1}
\begin{algorithmic}[1]
\While {$\Pi_{x,d}$ and $\Pi_{y,s}$ not converging}
\State Compute $\Pi_{x,d}(k+1)$  using \eqref{ADMM2_eqn1}, for all $x\in\mathcal{X}_y$
\State Compute $\Pi_{y,s}(k+1)$  using \eqref{ADMM2_eqn2}, for all $y\in\mathcal{Y}_x$
\State Compute $\pi_{xy}(k+1)$  using \eqref{ADMM2_eqn3}, for all $\{x,y\}\in \mathcal{E}$
\State Compute $\alpha_{xy}(k+1)$  using \eqref{ADMM2_eqn4}, for all $\{x,y\}\in \mathcal{E}$
\EndWhile
\State \textbf{return} $\pi_{xy}(k+1)$, for all $\{x,y\}\in \mathcal{E}$
\end{algorithmic}
\end{algorithm}

\section{Dynamic Distributed Optimal Transport}
In this section, we adapt the above distributed OT framework for applications to the UAV waypoint matching problem. For this application, consider the UAVs as the source and the set of waypoints as targets where each waypoint is a target node. 

\subsection{Parameter Adjustments} \label{subsec:param_adjust}
To fit Algorithm \ref{Alg:1} to the matching problem the constraints must be defined more rigidly. First, consider that each waypoint should only have one UAV at it at any given time, thus the constraint $\underline{q}_y \leq \sum_{x \in \mathcal{X}_y} \pi_{xy,s} \leq \bar{q}_y$ becomes $0 \leq \sum_{x \in \mathcal{X}_y} \pi_{xy,s} \leq 1$. On the source side, every UAV must go to a waypoint thus $\underline{p}_x \leq \sum_{y \in \mathcal{Y}_x} \pi_{xy,d} \leq \bar{p}_x$ becomes $\sum_{y \in \mathcal{Y}_x} = 1$. 

We also consider the network under linear parameters, making $d_{xy}(\pi_{xy,d})=\gamma_{xy}\pi_{xy,d}$ and $s_{xy}(\pi_{xy,s})=\delta_{xy}\pi_{xy,s}$. These parameters satisfy Assumption \ref{assump1} and are more applicable to the parameters we see in a UAV to waypoint matching problem.


\subsection{Dynamic Optimal Transport} \label{subsec:onlineOT}
The network structure and parameters need to be updated often; for example, every time a UAV reaches a the waypoint, the network parameters need to be updated so that waypoint is not visited by any agents in the swarm again. Also, the parameters of the nodes need to be updated based on the information gained at the waypoint. Other events like a UAV needing to land for a battery replacement can also affect the network. We develop a dynamic algorithm by adding in a time step to account for this. With the dynamic algorithm, the UAV can constantly be iterating, and when parameters are updated, it will converge to a new optimal solution without having to start a new set of iterations. Once the algorithm has converged, the matching scheme can be captured, and the UAV will have its next waypoint. We introduce a new set $t \in \mathcal{T} = \{1,2,...,T\}$ that indicates a time $t$. From the modifications in subsections \ref{subsec:param_adjust} and \ref{subsec:onlineOT}, we obtain the following iterative steps:
\begin{equation}\label{ADMM3_eqn1}
\begin{split}
    \Pi_{x,d}(k+1) &\in \arg \min_{\Pi_{x,d}\in\mathcal{F}_{x,d}} - \sum_{t \in \mathcal{T}}\sum_{y\in\mathcal{Y}_x} \gamma_{xy}\pi_{xy,d}^t
    + \sum_{t\in\mathcal{T}}\sum_{y\in\mathcal{Y}_x} \\&\alpha_{xy}(k) \pi_{xy,d}^t +\frac{\eta}{2} \sum_{t\in\mathcal{T}}\sum_{y\in\mathcal{Y}_x} \left(\pi_{xy,d}^t - \pi_{xy}^t(k)\right)^2,
\end{split}
\end{equation}
\begin{equation}\label{ADMM3_eqn2}
\begin{split}
    \Pi_{y,s}(k+&1) \in \arg \min_{\Pi_{y,s}\in\mathcal{F}_{y,s}} - \sum_{t \in\mathcal{T}}\sum_{x\in\mathcal{X}_y} \delta_{xy}\pi_{xy,s}^t -\sum_{t \in\mathcal{T}}\sum_{x\in\mathcal{X}_y}\\& \alpha_{xy}(k)\pi_{xy,s}^t + \frac{\eta}{2} \sum_{t \in\mathcal{T}}\sum_{x\in\mathcal{X}_y} \left(\pi_{xy}^t(k) - \pi_{xy,s}^t\right)^2
\end{split}
\end{equation}
\begin{equation}\label{ADMM3_eqn3}
\begin{split}
    \pi_{xy}^t(k+1) = \frac{1}{2} \left(\pi_{xy,d}^t(k+1) + \pi_{xy,s}^t(k+1)\right),
\end{split}
\end{equation}
\begin{equation}\label{ADMM3_eqn4}
\begin{split}
    \alpha_{xy}(k+1) = \alpha_{xy}(k) + \frac{\eta}{2}\left(\pi_{xy,d}^t(k+1) - \pi_{xy,s}^t(k+1)\right),
\end{split}
\end{equation}
\textit{where $\Pi_{\tilde{x},t}:=\{\pi_{xy,d}\}_{y\in\mathcal{Y}_x,x=\tilde{x},t\in\mathcal{T}}$ represents the solution at target node $\tilde{x}\in\mathcal{X}$, and $\Pi_{\tilde{y},s}:=\{\pi_{xy,s}\}_{x\in\mathcal{X}_y,y=\tilde{y},t\in\mathcal{T}}$ represents the proposed solution at source node $\tilde{y}\in\mathcal{Y}$. In addition, $\mathcal{F}_{x,d} := \{ \Pi_{x,d} | \pi_{xy,d} \geq 0, y\in\mathcal{Y}_x, \sum_{y \in \mathcal{Y}_x} \pi_{xy,d}^t = 1\}$, and $\mathcal{F}_{y,s} := \{ \Pi_{y,s} | \pi_{xy,s} \geq 0, x\in\mathcal{X}_y, 0 \leq \sum_{x \in \mathcal{X}_y} \pi_{xy,s} \leq 1 \}$} 
\begin{figure}[t]
	\centering
	\includegraphics[width=1\linewidth]{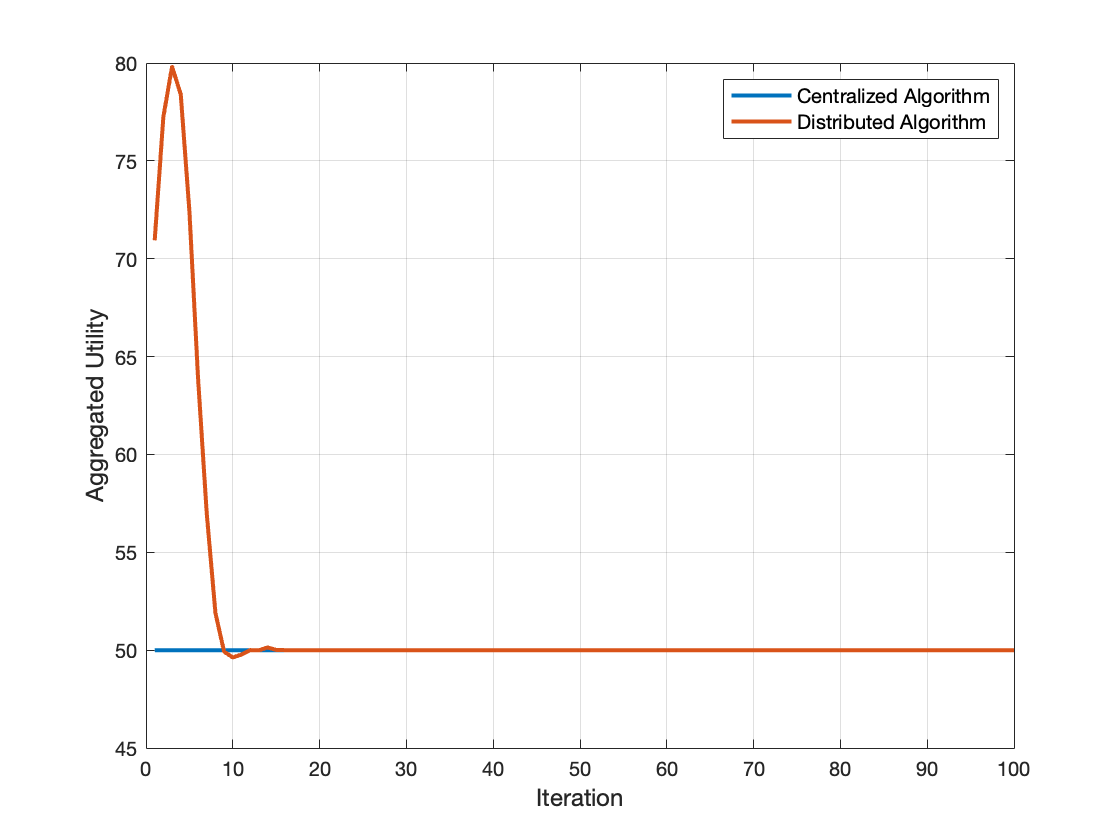}
	\caption{The distributed algorithm converges to the same solution of the centralized algorithm by showing the aggregation of the utility parameters in the network.}
	\label{fig:convergence}
\end{figure}


Once the iterative steps have converged, the UAV will communicate what waypoint it is heading to. When it has reached its calculated waypoint, it can take a reading from sensors or preform a required task and communicate the information from that point to the other UAVs. The other agents can then update the parameters $\delta_{xy}$ connecting to the waypoint node. Once the update is made, the iterative steps of the algorithm will begin to converge again. We also note that the iterative steps of the algorithm will not stop iterating. This ensures that a new optimal solution is found even if something out of the ordinary occurs. The compromised UAV can communicate this and the other UAVs can update their parameters and converge to a new solution rather than having to wait until they have arrived at their waypoint to find out that another UAV has dropped out. Also, the agents will most likely reach their waypoints asynchronously. With a dynamic algorithm, the UAVs will not need to wait for the other agents to reach their destinations to calculate their next waypoint. Instead, they can calculate the optimal waypoint for themselves at that time with the given information. For convenience, this is  summarized in Algorithm \ref{Alg:2}.
\begin{algorithm}[b]
\caption{Dynamic Distributed OT Algorithm for UAVs}\label{Alg:2}
\begin{algorithmic}[1]
\While {UAV is in the air }
\State Compute $\Pi_{x,d}(k+1)$  using \eqref{ADMM3_eqn1}, for all $x\in\mathcal{X}_y$
\State Compute $\Pi_{y,s}(k+1)$  using \eqref{ADMM3_eqn2}, for all $y\in\mathcal{Y}_x$
\State Compute $\pi_{xy}(k+1)$  using \eqref{ADMM3_eqn3}, for all $\{x,y\}\in \mathcal{E}$
\State Compute $\alpha_{xy}(k+1)$  using \eqref{ADMM3_eqn4}, for all $\{x,y\}\in \mathcal{E}$
\If{Convergence is reached}
    \State Interpret waypoint from $\pi_{xy}(k)$ 
    \State Go to waypoint
    \State Communicate next waypoint to swarm
\EndIf
\If{At waypoint}
    \State Update $\gamma_{xy}$ and $\delta_{xy}$
    \State Update network structure
    \State Communicate information at waypoint
\EndIf
\EndWhile
\end{algorithmic}
\end{algorithm}

\section{Preliminary Case Study}

In this section, we corroborate our algorithm by showing its convergence and how it considers the parameters of the UAVs and waypoints. For example, consider a swarm with three UAVs and ten waypoints and randomly generated linear parameters between 1 and 10 for both $\gamma_{xy}$ and $\delta_{xy}$:
\begin{equation*}
\begin{aligned}
\delta_{xy} = 
\begin{bmatrix}
    5 & 4 & 2 & 6 & 3 & 7 & 2 & 10 & 9 & 1 \\ 8 & 2 & 4 & 5 & 9 & 5 & 2 & 4 & 9 & 2 \\
    1 & 1 & 4 & 7 & 1 & 6 & 9 & 7 & 1 & 9
\end{bmatrix}
\end{aligned}
\end{equation*}
\begin{equation*}
\begin{aligned}
\gamma_{xy} = 
\begin{bmatrix}
    5 & 9 & 5 & 2 & 4 & 9 & 2 & 5 & 7 & 9 \\ 7 & 1 & 6 & 9 & 7 & 1 & 9 & 10 & 4 & 1 \\
    3 & 7 & 2 & 10 & 9 & 1 & 1 & 6 & 7 & 8
\end{bmatrix}
\end{aligned}
\end{equation*}

We first show that the distributed algorithm converges to the solution of a centralized algorithm. The convergence factor is set relatively high with $\eta=10$, this allows the algorithm to converge more quickly, and time is saved by shortening the number of iterations. The convergence of the distributed algorithm in Alg. \ref{Alg:1} to the centralized algorithm in \eqref{eqn:centralOT2} is shown in Fig \ref{fig:convergence}. 

Once the algorithm has converged it will output a 3-by-10 matrix of zeros and ones. The indices of the ones in the matrix indicate which UAV should go to which waypoint. In this case, UAV 1 will be sent to waypoint 8, UAV 2 will be sent to waypoint 5 and UAV 3 will be sent to waypoint 10.

We can corroborate the output of the algorithm by examining the aggregated utility of the parameters $d_{xy}$ and $s_{xy}$.
\begin{equation*}
\delta_{xy} + \gamma_{xy}=
    \begin{bmatrix}
        6 & 10 & 9 & 14 & 6 & 12 & 5 & 17 & 14 & 3 \\ 13 & 9 & 13 & 15 & 17 & 15 & 4 & 7 & 10 & 8 \\ 11 & 5 & 5 & 15 & 3 & 9 & 10 & 10 & 7 & 16
    \end{bmatrix}
\end{equation*}
As this example is relatively simple, we can easily see that the algorithm is sending the UAV to the waypoint with the highest utility as we expect it do. 

\subsection{Dynamic Algorithm}
\begin{figure}[!t]
    \centering
    \includegraphics[width=.95\columnwidth]{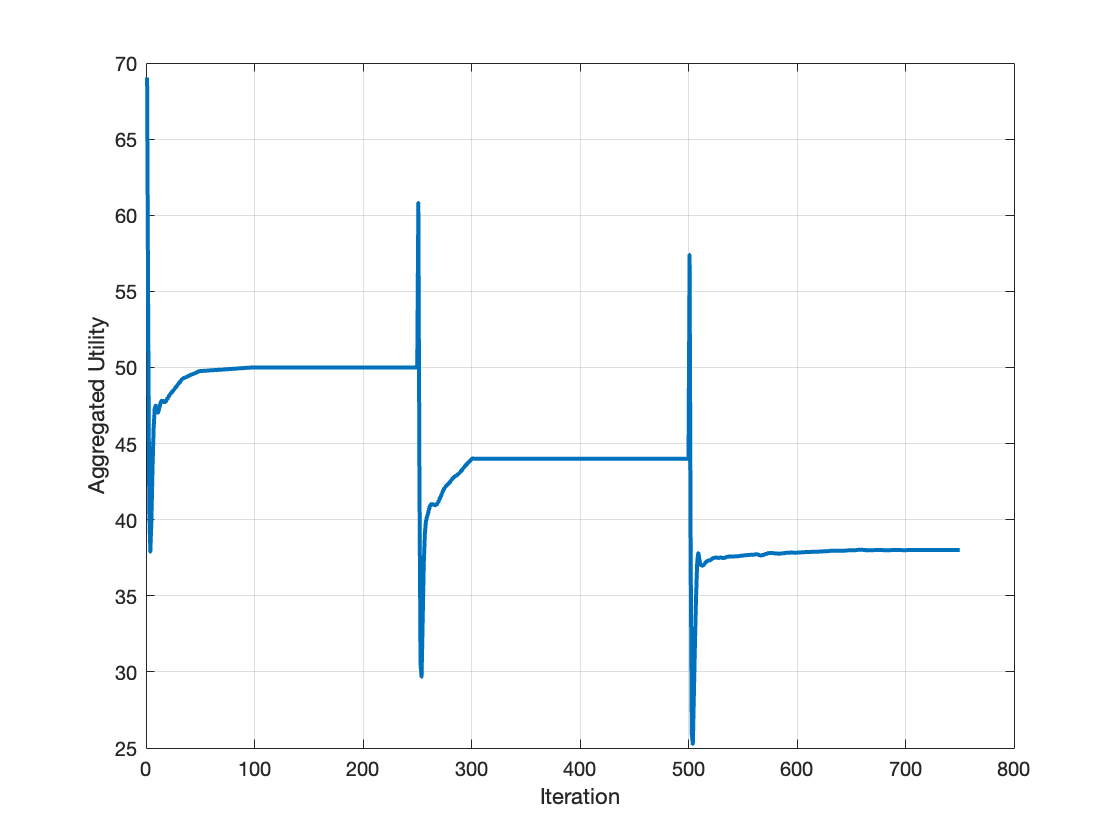}
    \caption{The online algorithm converges to the new optimal solution even when parameters are adjusted while iterating.}
    \label{fig:online}
\end{figure}

Next we verify the results of the online algorithm. In this study, at each iteration the network is updated to exclude the waypoints that have already been visited, this ensures that the waypoint will not be visited twice by any of the UAVs. The parameters $\gamma_{xy}$ and $\delta_{xy}$ are also updated to capture the new updated distances and importance of the surrounding waypoints. For the parameter updates we randomly generate new parameters for this study. These updates to the network are made after every 250 iterations, but the updates to the parameters or network structure can be made at any iteration. The algorithm converges to the new optimal solution when the parameters are updated, as shown in Fig. \ref{fig:online}

The online algorithm is advantageous for this application because the network will update often. Every time a UAV reaches a waypoint it will communicate sensor readings and location to the other agents which requires the network to update so that other UAVs can calculate their optimal waypoint. As shown in Fig. \ref{fig:online} the algorithm converges in about 50 iterations, so updates can be made often and a new optimal solution can be computed with relative ease. 

\section{Simulation}
\begin{figure*}[!t] 
\centering
\subfigure[Waypoint Map]{\includegraphics[width=0.69\columnwidth]{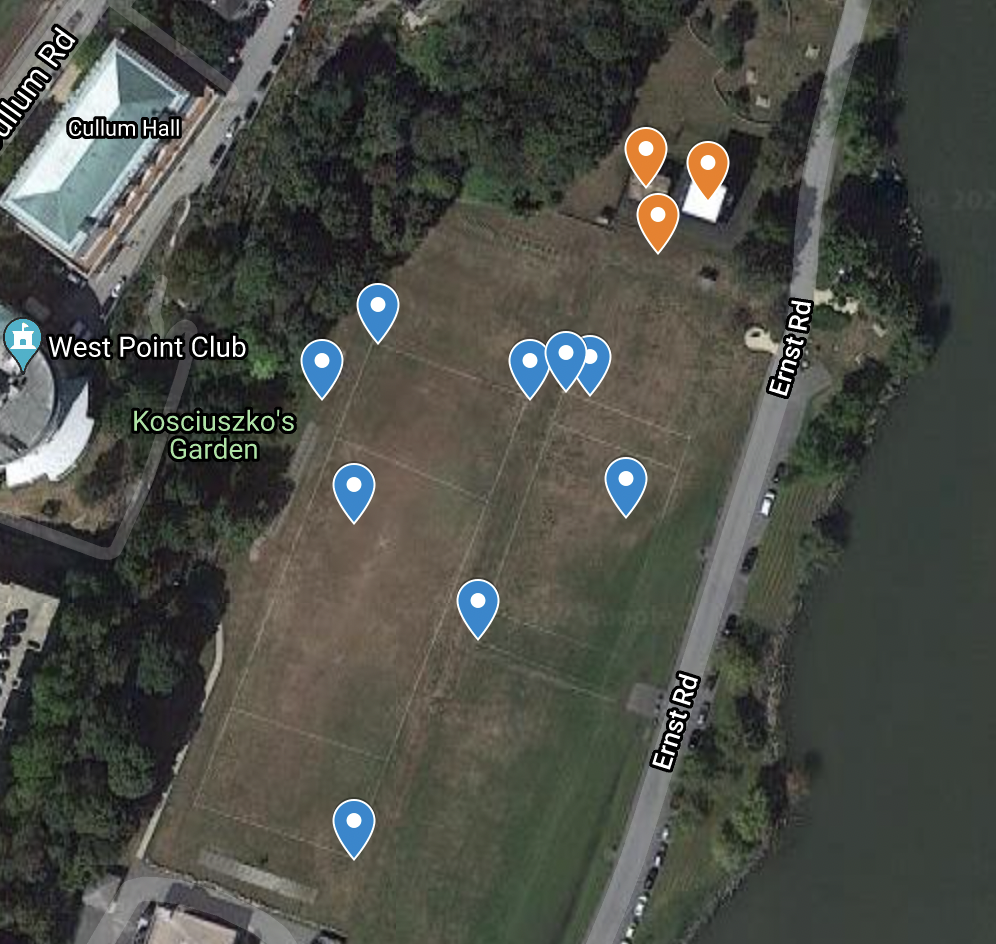}\label{fig:wp_map}}
\subfigure[Greedy Allocation]{\includegraphics[width=0.67\columnwidth]{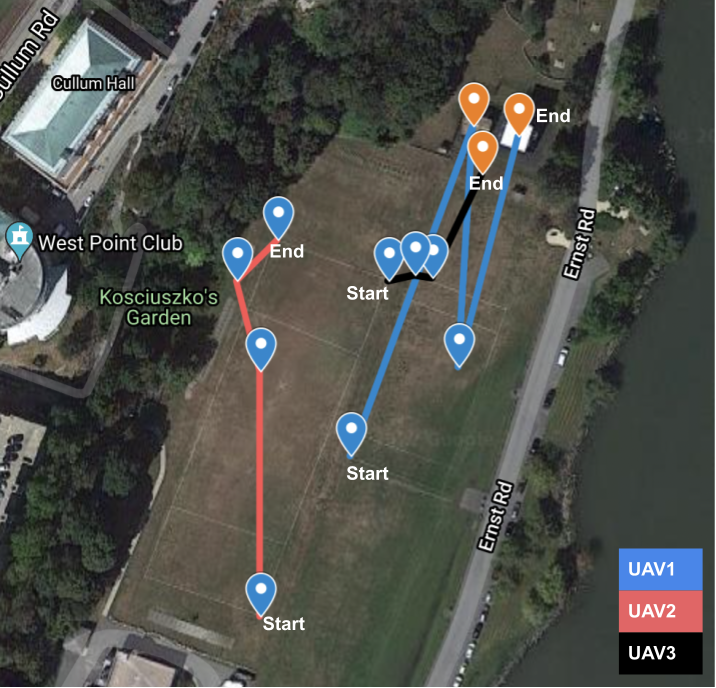}\label{fig:greedy}}
\subfigure[Optimal Transport Algorithm ]{\includegraphics[width=0.66\columnwidth]{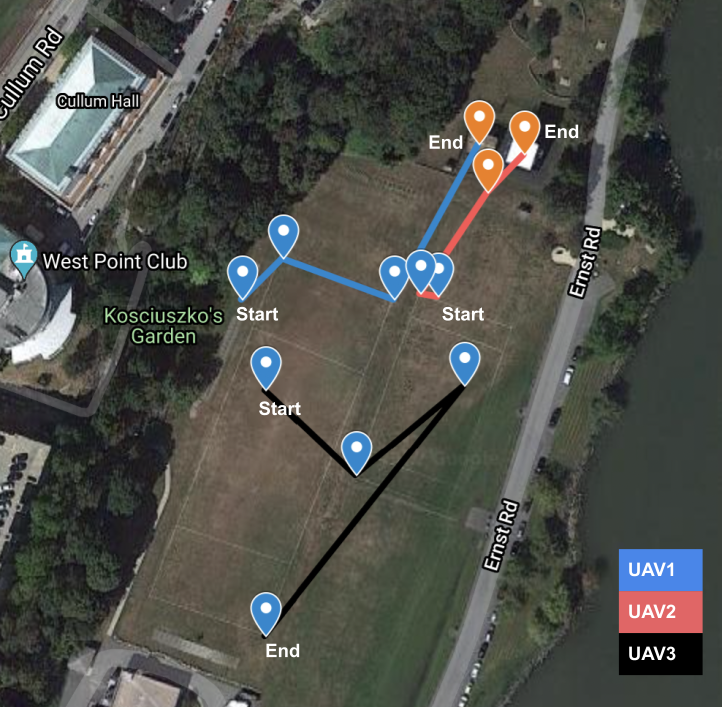}\label{fig:optimal_map}}
\caption{(a) shows the location of the twelve waypoints, blue indicates a waypoint where there is no chemical present, orange indicates a waypoint with chemical, (b) shows the path of the UAVs with the greedy algorithm, (c) shows the path of the UAVs using Alg.\ref{Alg:2}}
\label{fig:maps}
\end{figure*}
In this section we compare the algorithm outline in Alg. \ref{Alg:2} to a greedy assignment algorithm in simulation. We simulate a swarm of UAVs with a chemical sensor that will take a reading at multiple waypoints to find where the chemical is strongest. We assume that the UAVs are altitude deconflicted in this exercise. 
\subsection{Setup}
Consider an environment with three UAVs and twelve waypoints to be scanned, three of those waypoints are considered to have a chemical. The actual chemical sensor records a numerical reading, in simulation we report a binary reading, meaning there is chemical present or there is not, for simplicity. The waypoints correspond to an area with a field and a few small buildings and the area around the buildings is the area where the chemical is considered present, as shown in Fig. \ref{fig:wp_map}. 

We compare our algorithm to a greedy assignment algorithm. The greedy algorithm assigns four waypoints to each UAVs before they start. It assigns waypoints solely based on distance. The point that is closest and not taken by another UAV already will be assigned to that agent. While this algorithm does consider distance it does not consider it at run time so a UAV maybe a assigned the closest waypoint without the consideration of the other UAVs. The algorithm is greedy because the agent that comes online first will choose the waypoints that are best for itself without considering the other UAVs. Thus a waypoint may be assigned to one UAV even if it is far closer to another UAV. 

For the optimal transport algorithm in Alg. \ref{Alg:2}, the UAVs need to be assigned their first waypoint since they all take off at the same location. The first waypoint is assigned such that UAV one goes to waypoint one, UAV two goes to waypoint two, and so on. We do this since the distances to the waypoints will be the same and there is nothing known about the chemical sensor readings at the waypoints.

\subsection{Results}
The allocation schemes of the greedy and dynamic distributed optimal transport algorithms are show in Fig. \ref{fig:greedy} and Fig. \ref{fig:optimal_map} respectively. Table \ref{table:datances} shows the distances each UAV traveled between waypoints in meters.  

\begin{table}[h!]
\centering
\begin{tabular}{||c| c| c||} 
 \hline
 \textbf{UAV} & \textbf{Greedy} &  \textbf{OT} \\  
 \hline\hline
 1 & 255m & 108m \\ 
 2 & 122m & 59m \\
 3 & 49m & 179m  \\
 \textbf{Total} & 426m & 346m\\
 \hline
\end{tabular}
\caption{Distances traveled}
\label{table:datances}
\end{table}
The greedy algorithm has a total distance traveled of 426 meters while the optimal transport algorithm has a total distance traveled of 346 meters. The smaller travel distance is essential for preserving battery life. We also highlight that the distance are more varied even among the greedy algorithm than with the optimal transport algorithm. The greedy algorithm has UAV one traveling over double the distance of the other UAVs using more of that batteries life to scan the same number of waypoints as the other agents. The lesser total distance of the OT algorithm means that the scan of  the waypoints will be completed quicker, saving time. We note that the algorithm does take time to converge in practice, in simulation it take around two seconds to converge but realize this could be longer on the constrained hardware of a UAV.   These metrics show the effectiveness of the algorithm and provides a better way to allocate the waypoints to each agent in a swarm. 

\section{Conclusion \& Future Work}
In this paper we developed a dynamic and distributed algorithm for the efficient matching of UAV swarm agents to waypoints. The algorithm is capable of considering a parameter on both the agent and waypoint side and can converge to a new optimal matching scheme when parameters in the network change such as a UAV needing to land for a new battery or updating parameters based on sensor readings. The simulation shows the effectiveness of this algorithm and showed that the developed algorithm was able to find a more efficient way of navigating the waypoints than a greedy assignment algorithm by cutting down the total distance that the UAVs needed to travel thus saving battery life and shortening mission time. 

Future work includes the developing a similar algorithm for a more complex use case such as a heterogeneous swarm of robots. We also plan to cut down on time to reach convergence making the algorithm better suited for actual hardware. We also consider extending the security measures of the swarm especially looking into communication between the swarm with a MPU 5 or equivalent, extending or combining simulation with other swarm missions and building effective yet flexible templates for mission commanders to support their specific parameters. 

\section{Acknowledgements}
This research was sponsored by the Army Research Laboratory and was accomplished under Cooperative Agreement Number W911NF-21-2-0281. The views and conclusions contained in this document are those of the authors and should not be interpreted as representing the official policies, either expressed or implied, of the Army Research Laboratory or the U.S. Government. The U.S. Government is authorized to reproduce and distribute reprints for Government purposes notwithstanding any copyright notation herein.
\bibliographystyle{IEEEtran}
\bibliography{references.bib}

\begin{thebibliography}{10}
\providecommand{\url}[1]{#1}
\csname url@samestyle\endcsname
\providecommand{\newblock}{\relax}
\providecommand{\bibinfo}[2]{#2}
\providecommand{\BIBentrySTDinterwordspacing}{\spaceskip=0pt\relax}
\providecommand{\BIBentryALTinterwordstretchfactor}{4}
\providecommand{\BIBentryALTinterwordspacing}{\spaceskip=\fontdimen2\font plus
\BIBentryALTinterwordstretchfactor\fontdimen3\font minus
  \fontdimen4\font\relax}
\providecommand{\BIBforeignlanguage}[2]{{%
\expandafter\ifx\csname l@#1\endcsname\relax
\typeout{** WARNING: IEEEtran.bst: No hyphenation pattern has been}%
\typeout{** loaded for the language `#1'. Using the pattern for}%
\typeout{** the default language instead.}%
\else
\language=\csname l@#1\endcsname
\fi
#2}}
\providecommand{\BIBdecl}{\relax}
\BIBdecl

\bibitem{sanders2017}
A.~Sanders, ``Drone swarms,'' \emph{Masters Thessis}, 2017.

\bibitem{galichon2018optimal}
A.~Galichon, \emph{Optimal Transport Methods in Economics}.\hskip 1em plus
  0.5em minus 0.4em\relax Princeton University Press, 2018.

\bibitem{arnold2019robot}
R.~Arnold, K.~Carey, B.~Abruzzo, and C.~Korpela, ``What is a robot swarm: a
  definition for swarming robotics,'' in \emph{2019 IEEE 10th Annual Ubiquitous
  Computing, Electronics \& Mobile Communication Conference (UEMCON)}.\hskip
  1em plus 0.5em minus 0.4em\relax IEEE, 2019, pp. 0074--0081.

\bibitem{arnold2021performance}
R.~Arnold, E.~Mezzacappa, M.~Jablonski, J.~Jablonski, and B.~Abruzzo,
  ``Performance comparison of decentralized undirected swarms versus
  centralized directed swarms at different levels of quality of knowledge,'' in
  \emph{2021 IEEE International Symposium on Technologies for Homeland Security
  (HST)}.\hskip 1em plus 0.5em minus 0.4em\relax IEEE, 2021, pp. 1--9.

\bibitem{brick2018}
T.~Brick, M.~Lanham, A.~Kopeikin, C.~Korpela, and R.~Morales, ``Zero to swarm:
  Integrating suas swarming into a multi-disciplinary engineering program,'' in
  \emph{2018 International Conference on Unmanned Aircraft Systems (ICUAS)},
  2018, pp. 308--314.

\bibitem{sahin2005swarms}
E.~{\c{S}}ahin, ``Swarm robotics: From sources of inspiration to domains of
  application,'' in \emph{Swarm Robotics}, E.~{\c{S}}ahin and W.~M. Spears,
  Eds.\hskip 1em plus 0.5em minus 0.4em\relax Berlin, Heidelberg: Springer
  Berlin Heidelberg, 2005, pp. 10--20.

\bibitem{zhang2019consensus}
R.~Zhang and Q.~Zhu, ``Consensus-based distributed discrete optimal transport
  for decentralized resource matching,'' \emph{IEEE Transactions on Signal and
  Information Processing over Networks}, vol.~5, no.~3, pp. 511--524, 2019.

\bibitem{jhughes2021fair}
J.~Hughes and J.~Chen, ``Fair and distributed dynamic optimal transport for
  resource allocation over networks,'' in \emph{55th Annual Conference on
  Information Sciences and Systems (CISS)}, 2021.

\bibitem{behcet2014probabilisticOT}
B.~Açıkmeşe and D.~S. Bayard, ``Probabilistic swarm guidance for
  collaborative autonomous agents,'' in \emph{2014 American Control
  Conference}, 2014, pp. 477--482.

\bibitem{vishaal2018dot}
V.~Krishnan and S.~Martínez, ``Distributed optimal transport for the
  deployment of swarms,'' in \emph{2018 IEEE Conference on Decision and Control
  (CDC)}, 2018, pp. 4583--4588.

\bibitem{wang2020dynamic}
\BIBentryALTinterwordspacing
Q.~Wang and X.~Mao, ``Dynamic task allocation method of swarm robots based on
  optimal mass transport theory,'' \emph{Symmetry}, vol.~12, no.~10, 2020.
  [Online]. Available: \url{https://www.mdpi.com/2073-8994/12/10/1682}
\BIBentrySTDinterwordspacing

\bibitem{savidge2019radiationswarm}
B.~Savidge, A.~Kopeikin, R.~Arnold, and D.~Larkin, ``Uas swarm shares survey
  data to expedite coordinated mapping of radiation hotspots,'' in \emph{2019
  IEEE International Symposium on Technologies for Homeland Security (HST)},
  2019, pp. 1--7.

\bibitem{kopeiken2019swarmdata}
\BIBentryALTinterwordspacing
A.~Kopeikin, S.~Heider, D.~Larkin, C.~Korpela, R.~Morales, and J.~E. Bluman,
  \emph{Unmanned Aircraft System Swarm for Radiological and Imagery Data
  Collection}. [Online]. Available:
  \url{https://arc.aiaa.org/doi/abs/10.2514/6.2019-2286}
\BIBentrySTDinterwordspacing

\bibitem{Ghorbanzadeh2017}
\BIBentryALTinterwordspacing
M.~Ghorbanzadeh, A.~Abdelhadi, and C.~Clancy, \emph{Distributed Resource
  Allocation}.\hskip 1em plus 0.5em minus 0.4em\relax Cham: Springer
  International Publishing, 2017, pp. 61--91. [Online]. Available:
  \url{https://doi.org/10.1007/978-3-319-46267-7_4}
\BIBentrySTDinterwordspacing

\bibitem{Niu2013distributedRA}
D.~Niu and B.~Li, ``An efficient distributed algorithm for resource allocation
  in large-scale coupled systems,'' in \emph{2013 Proceedings IEEE INFOCOM},
  2013, pp. 1501--1509.

\bibitem{Schmidt2009distrivutedSchemes}
D.~A. Schmidt, C.~Shi, R.~A. Berry, M.~L. Honig, and W.~Utschick, ``Distributed
  resource allocation schemes,'' \emph{IEEE Signal Processing Magazine},
  vol.~26, no.~5, pp. 53--63, 2009.

\bibitem{boyd2011distributed}
S.~Boyd, N.~Parikh, and E.~Chu, \emph{Distributed optimization and statistical
  learning via the alternating direction method of multipliers}.\hskip 1em plus
  0.5em minus 0.4em\relax Now Publishers Inc, 2011.

\end{thebibliography}

\end{document}